\documentclass[twocolumn,showpacs,preprintnumbers,amsmath,amssymb]{revtex4}
\usepackage{graphicx}

\begin{document}
\title{\bf Casimir Energy for a Coupled Fermion-Kink System and its stability}
\author{S.S. Gousheh}
\email{ss-gousheh@sbu.ac.ir}
\author{A. Mohammadi}
\email{a_ mohammadi@sbu.ac.ir}
\author{L. Shahkarami}
\email{l_shahkarami@sbu.ac.ir}

\affiliation{%
Department of Physics, Shahid Beheshti University G.C., Evin, Tehran
19839, Iran
}%
\date{\today}

\begin{abstract}
We compute the Casimir energy for a system consisting of a fermion and a pseudoscalar field in the form of a prescribed kink. This model is not exactly solvable and we use the phase shift method to compute the Casimir energy. We use the relaxation method to find the bound states and the Runge-Kutta-Fehlberg method to obtain the scattering wavefunctions of the fermion in the whole interval of $x$. The resulting phase shifts are consistent with the weak and strong forms of the Levinson theorem. Then, we compute and plot the Casimir energy as a function of the parameters of the pseudoscalar field, i.e. the slope of $\phi(x)$ at $x=0$ ($\mu$) and the value of $\phi(x)$ at infinity ($\theta_0$). In the graph of the Casimir energy as a function of $\mu$ there is a sharp maximum occurring when the fermion bound state energy crosses the line of $E=0$. Furthermore, this graph shows that the Casimir energy goes to zero for $\mu\rightarrow 0$, and also for $\mu\rightarrow \infty$ when $\theta_0$ is an integer multiple of $\pi$. Moreover, the graph of the Casimir energy as a function of $\theta_0$ shows that this energy is on the average an increasing function of $\theta_0$ and has a cusp whenever there is a zero fermionic mode. We finally compute the total energy of a system consisting of a valence fermion in the ground state. Most importantly, we show that this energy (the sum of the Casimir energy and the energy of the fermion) is minimum when the background field has winding number one, independent of the details of the background profile. Throughout the paper we compare our results with those of a simple exactly solvable model, where a piece-wise linear profile approximates the kink. We find that the kink is an almost reflectionless barrier for the fermions, within the context of our model.
\end{abstract}
\maketitle

\section{Introduction}
The Casimir effect arises from the distortion of the zero-point energy of a quantum field due to the presence of non-trivial background fields or the imposition of non-trivial boundary conditions. This effect was first proposed by Casimir in 1948 \cite{casimir1,casimir2}, when he predicted the existence of an attractive force between two neutral infinite parallel metallic plates in a vacuum, placed a few micrometers apart. Since Casimir's work, this effect has attracted much interest and many authors have calculated the Casimir energy and the resulting force caused by the presence of non-trivial boundary conditions for various geometries such as parallel plates, cylinders, and spheres \cite{casimir1,plate1,plate2,plate3,plate4,plate5,deraadcylinder,cylinder1,cylinder2,cylinder3,cylinder4,cylinder5,boyersphere,sphere1,sphere2,sphere3,sphere4} and other geometries \cite{othergeometry1,othergeometry2,othergeometry3,othergeometry4,othergeometry5,othergeometry6,othergeometry7,othergeometry8}. Moreover, they have used many different regularization and analytic continuation schemes to remove the divergences. Some of these techniques are the heat-kernel method \cite{heat1,heat2}, the Green function
formalism \cite{green}, the mode number summation method combined with the
zeta function analytic continuation technique
\cite{modenumber1,modenumber2,modenumber3,modenumber4}, and the
multiple scattering expansions \cite{scatter}. The first experimental attempt to observe this phenomenon was conducted by Marcus Sparnaay \cite{sparnaay} in 1958. In this experiment two parallel metallic plates were used, and the results had a very poor accuracy. In 1997, Steve K. Lamoreaux
\cite{Lamorea1,Lamorea2} measured the Casimir energy with a high accuracy using a plate and a metallic spherical shell, and this was the first successful experiment to verify the Casimir effect. Since then, many different experiments have been performed to measure the Casimir energy for various geometries \cite{exper1,exper2,exper3,exper4,exper5,exper6,exper7}.
 \par As mentioned above, the zero-point energy can also be affected by the presence of non-trivial background fields. The background field is usually chosen to be a soliton. Also sometimes a very simple potential such as an electric potential well is chosen as the background field. This simple choice renders the problems of vacuum polarization and the Casimir energy exactly solvable \cite{dehghan}. The Casimir energy also contributes to the lowest order quantum correction to the mass of the soliton. Many authors use this correction for models containing solitons such as supersymmetric solitons
 \cite{dashen,solitonmass1,solitonmass2,solitonmass3,solitonmass4,solitonmass5,solitonmass6,solitonmass7,solitonmass8,heavy1,heavy2,heavy3,heavy4,super1,super2,super3,la}. For most of the models with solitons, the problem is not exactly solvable. Sometimes indirect methods such as the phase shift method which relates the derivative of the phase shift with respect to the momentum to the spectral deficiency in the continuum states are used to calculate the Casimir energy \cite{dashen,heavy3,heavy4,super1,dehghan}.
\par In this paper we calculate the Casimir energy for a system containing a Fermi field chirally coupled to a pseudoscalar field which is prescribed and has the form of an isolated kink. In a previous work \cite{la} we calculated the Casimir energy for a similar system where the soliton profile is approximated by a piece-wise linear function, and this renders the problem exactly solvable. The vacuum polarization for that model has also been calculated \cite{dr}. For that problem we calculated the Casimir energy exactly and directly by subtracting the vacuum energy of the system in the absence from the presence of the disturbance which is the pseudoscalar field. Throughout this paper we shall refer to that model as the simple exactly solvable model. However, the present model is not analytically solvable and we use the phase shift method to compute the Casimir energy. As usually happens, the presence of the disturbance, e.g. the kink, leads to the appearance of one or more discrete bound states and also changes the continuum wavefunctions as compared to the free case. These changes have many manifestations including induced vacuum polarization and Casimir energy of the system. We have previously investigated a similar system where neither the Fermi field nor the pseudoscalar field, with boundary values of a topologically non-trivial configuration, were prescribed. They were allowed to interact and the non-perturbative final results, i.e. the results beyond the first order ``back reaction" ones, revealed that the actual solitary wave profile differs only very slightly from an isolated kink \cite{leila}. This proximity is one of our motivations to study the properties of the coupled fermion-kink system.
\par One of the main purposes of this paper is to investigate how the functional form of a pseudoscalar background field ($\phi(x)$) affects the properties of the coupled fermion-pseudoscalar field system. Some of these effects have already been investigated (see for example \cite{jackiw}). For example it is well known that changing the value of $\phi(x)$ at spatial infinity ($\pm \theta_0$) affects the spectral deficiency in the continua, and this usually leads to what is called the adiabatic contribution to the induced vacuum polarization \cite{goldstone,mackenzie}. Moreover, changing the spatial profile of the background field close to the center e.g. the value of the slope of $\phi(x)$ at $x=0$ ($\mu$), changes the pattern of energy levels crossing $E=0$. These crossings lead to what is usually called the non-adiabatic contribution to the vacuum polarization \cite{mackenzie,dr}. The same changes in the spectrum of the system that lead to the induced vacuum polarization, also affect its Casimir energy. In this paper we investigate the effects of the functional form of $\phi(x)$ on the Casimir energy and stability of the system. In order to accomplish this we meticulously study and compare the properties of the system which has the kink as the background field with those of the simple exactly solvable model. For the comparison to be meaningful and refined, we choose both background fields to have the same $\theta_0$ and $\mu$. In other words, this comparison serves a dual purpose: we not only investigate the behavior of the systems as a function of the parameters $\theta_0$ and $\mu$, but also investigate the difference between these two systems for the same parameters. The latter investigation allows us to explore the effect of the finer details of $\phi(x)$ on the overall properties of the system. 
\par In section 2 we briefly explain how to find the continuum scattering wavefunctions of this system using the Runge-Kutta-Fehlberg method of order 6. Then, in section 3 we compute the phase shifts. Using the relation between the phase shift and the difference between the density of states in the presence and absence of the disturbance, we can write an expression for the Casimir energy in terms of the phase shift. We then calculate and plot the Casimir energy as a function of the parameters of the pseudoscalar field, i.e. the slope of the pseudoscalar field $\phi(x)$ at $x=0$ ($\mu$) and the value of the field at infinity ($\theta_0$). In both cases we also show the results for the simple exactly solvable model, for comparison. In section 4 we add the Casimir energy to the energy of a system consisting of a valence fermion in the ground state, when there is a soliton as the background field and discuss the stability of the system. For each result obtained and displayed for our model, we also present the corresponding results of the simple exactly solvable model, for comparison. In section 5 we summarize and discuss our conclusions.

\section{The spectrum of the fermion in the presence of the prescribed kink}
\begin{center}
\begin{figure}[th] \includegraphics[width=6.9cm]{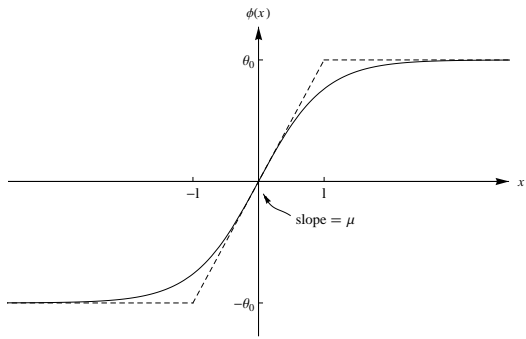}\caption{\label{kinkdr} \small
   The solid and dashed lines show $\phi(x)$ for the kink and the simple exactly solvable model, respectively. The parameters $\theta_0$ and $\mu$ are also shown in the figure.}
  \label{geometry}
\end{figure}
\end{center}

We consider the coupling of a Fermi field and a pseudoscalar field governed by the following Lagrangian
\begin{equation}\label{e1}\vspace{.2cm}
 {\cal L}=\bar{\psi}\left(i\gamma^{\mu}\partial_{\mu}-M \mathrm{e}^{i \phi(x)\gamma^5}\right)\psi,
\end{equation}
and we choose $\phi(x)$ to be prescribed in the form of $\phi(x)=m/\sqrt{\lambda}\tanh \left[m x/\sqrt{2}\right]$ which is an isolated kink. The parameters $M$ and $m$ denote the mass of the Fermi and pseudoscalar field, respectively. Our purpose is to calculate the Casimir energy of this system. For exactly solvable systems we usually compute the complete spectrum of the fermion, including the bound states with their discrete energies and the continuum states, and then we can calculate the Casimir energy directly by subtracting the vacuum energy of the system in the presence and absence of the disturbance. However, the form chosen for the $\phi(x)$ makes the Euler-Lagrange equation of $\psi$ analytically unsolvable. Therefore, we have to use an appropriate numerical method to find the fermion spectrum. In order to facilitate the numerical calculations, we take advantage of the solutions of the exactly solvable model for choosing the initial values as well as comparison purposes. The Lagrangian for the simple exactly solvable model mentioned earlier is also the one shown in Eq.\,(\ref{e1}). However, the form chosen for $\phi(x)$ is a piece-wise linear which renders the problem exactly solvable. The $\phi(x)$ for this model is as follows
\begin{equation}\label{e8}
\phi(x)=\begin{cases}
-\theta_0& \mathrm{for } \hspace{.3cm}x \leqslant -l ,\\
\mu x& \mathrm{for }\hspace{.3cm} -l\leqslant x\leqslant l,\\
+\theta_0 & \mathrm{for }\hspace{.3cm} l \leqslant x.
\end{cases}
\end{equation}
This form for the pseudoscalar field along with the kink is shown in Fig.\,(\ref{kinkdr}). In this figure we indicate the parameters $\mu$, the slope of $\phi(x)$ at $x=0$, and $\pm \theta_0$, its values at the boundaries.
\subsection{The bound state energies and wavefunctions}
The spectrum of a Dirac field coupled to a background field gets distorted as compared to the free case. These distortions can be observed as spectral deficiencies in the continua and also the bound states could appear due to the presence of the background field.
\par Choosing the representation $\gamma^0=\sigma_1$, $\gamma^1=i\sigma_3$ and $\gamma^5=\gamma^0\gamma^1=\sigma_2$ for the Dirac matrices, the Dirac equation for the Lagrangian of Eq.\,(\ref{e1}) becomes
\begin{equation}\label{e3}
i\sigma_1\partial_t\psi - \sigma_3\partial_x\psi - M \left[\cos
\phi\left(x,t\right)+i\sigma_2 \sin \phi\left(x,t\right)\right]\psi
=0,
\end{equation}
where $\psi=\left(\! \begin{array}{c}
\psi_{1}\\
\psi_2
\end{array}\! \right)$. We define
\begin{equation}\label{e4}
\xi(x,t)=e^{-iEt} \left(\! \begin{array}{c}\xi_1(x)\\
\xi_2(x)\end{array}\! \right)=\left(\! \begin{array}{c}\psi_1+i\psi_2\\
\psi_1-i\psi_2\end{array}\! \right).
\end{equation}
The equation obeyed by $\xi(x,t)$ is
\begin{equation}\label{e5}
 \left(\! \begin{array}{cc}i\partial_x-E&\, i M \mathrm{e}^{i \phi(x)}\\
-i M \mathrm{e}^{-i \phi(x)}&\, -i\partial_x-E\end{array}\! \right) \left(\! \begin{array}{c}\xi_1\\
\xi_2\end{array}\! \right)= \left(\! \begin{array}{c}0\\
0\end{array}\! \right).
\end{equation}
In order to obtain the energies of the bound states of the fermion, we use a numerical method called the relaxation method (In an earlier paper \cite{leila} we have considered the bound states of this system in detail). This method is used for solving the boundary value problems. To solve $N$ real coupled first-order ODEs, we need $N$ boundary conditions, some of them are to be imposed at one boundary and the rest at the other boundary. The relaxation method determines the solution by starting with a guess and improving it, iteratively. We separate the real and imaginary parts of the upper and lower components of $\left(\! \begin{array}{c}\xi_1(x)\\
\xi_2(x)\end{array}\! \right)$ as $\xi_1(x)=y_1(x)+i y_2(x)$ and $\xi_2(x)=y_3(x)+i y_4(x)$. Therefore, the equations of motion for $y_i$s are as follows
\begin{align}\label{e6}
&y^{'}_1+M\cos \phi\left(x\right)y_3-E y_2-M\sin \phi\left(x\right)y_4=0,\\
&y^{'}_2+M\cos \phi\left(x\right)y_4+E y_1+M\sin
\phi\left(x\right)y_3=0,\\
&y^{'}_3+M\cos \phi\left(x\right)y_1+E y_4+M\sin
\phi\left(x\right)y_2=0,\\\label{e42} &y^{'}_4+M\cos
\phi\left(x\right)y_2-E y_3-M\sin \phi\left(x\right)y_1=0,
\end{align}
where prime denotes the derivative with respect to $x$. We have an additional (fifth) equation $E'=0$. To solve this set of five coupled first-order ODEs we need five conditions on the initial and final boundary points of the domain of the spatial variable, which could be chosen to be just $[0,\infty)$ instead of $(-\infty,+\infty)$, due to the invariance of the Lagrangian under the parity. Then, we map the $x$-interval $[0,\infty)$ to $[0,1]$ by the transformation $X=\tanh(x)$. Hence, the two boundaries of $X$ are $X=0$ and $X=1$. For bound states we use the relaxation method and choose the conditions at these boundaries as follows. At $X=0$ we choose two conditions: one parity condition and one assigning a value to one of the $y_i(0)$s. This value is allowed
to change so as to normalized $\psi(x)$. At $X = 1$ we choose three conditions: three of the $y_i(1)$s are set to zero. From now on we rescale all the quantities of the problem with respect to the mass of the Fermi field ($M$).
\par
\begin{center}
\begin{figure}[th] \includegraphics[width=6.6cm]{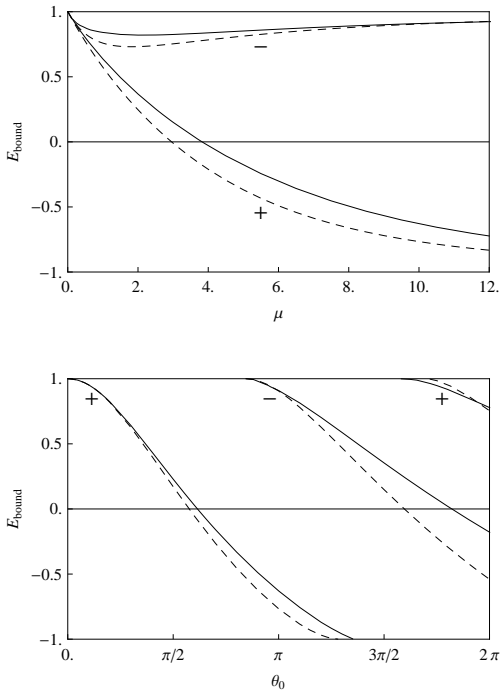}\caption{\label{bnd} \small
The energies of the bound states of the fermion. The upper graph shows the bound energies as a function of $\mu$ at $\theta_0=\pi$ and the lower graph shows the bound energies as a function of $\theta_0$ at $\mu=10$. In both graphs the solid and dashed lines are for the model with kink and the exactly solvable model, respectively. The parity of each bound state is indicated on the graphs by $\pm$ signs.}
  \label{geometry}
\end{figure}
\end{center}

 Some examples of the bound states obtained from numerical results are depicted in Fig.\,(\ref{bnd}). The upper graph shows the bound energy levels of the fermion as a function of $\mu$ at $\theta_0=\pi$, i.e. a soliton with winding number 1 and the lower graph shows the bound energies as a function of $\theta_0$ for the slope $\mu=10$. In both graphs the bound energy levels for the model with kink and the simple exactly solvable model are shown by solid and dashed lines, respectively. The $\pm$ signs refer to the parity of each of the bound states. Notice that the rather small difference between the profiles of these two background fields produces considerable difference between the pattern of bound state energies.
\subsection{The continuum scattering states}
 Now we focus on the continuum states. We can obtain the eigenstates of the Hamiltonian of the system, which can be chosen to be also the parity eigenstates. These solutions have to satisfy the parity condition $P\xi(x, t) =-\sigma_2\xi(-x, t)$. However, we obtain the wavefunctions which describe the scattering of a plane wave incident on the scattering region from the left, instead. The reason for obtaining the scattering solutions instead of the parity eigenstates is as follows. In the scattering process the wavefunctions on the right-hand side, far enough from the scattering region, have the $x$-dependence in the form of $\mathrm{e}^{i k x}$. Thus, by extracting the factor $\mathrm{e}^{i k x}$ from the solution, the remaining part of the solution on the far right would approach a constant, independent of the spatial variable and this will simplify the numerical analysis considerably. We choose the wavefunction for the scattering process to be in the form $$\xi_k(x)=\mathrm{e}^{i k x}\left(\! \begin{array}{c}\eta_1(x)+i\eta_2(x)\\
\eta_3(x)+i\eta_4(x)\end{array}\! \right),$$ where in general $\eta_i$s are real functions of $x$. Then, the equations of motion satisfied by $\eta_i$s would be in the following form
\begin{align}\label{e7}
&\eta^{'}_1+\cos \phi\left(x\right)\eta_3-(E+k) \eta_2-\sin \phi\left(x\right)\eta_4=0,\\\label{e71}&\eta^{'}_2+\cos \phi\left(x\right)\eta_4+(E+k) \eta_1+\sin
\phi\left(x\right)\eta_3=0,\\\label{e72}
&\eta^{'}_3+\cos \phi\left(x\right)\eta_1+(E-k) \eta_4+\sin
\phi\left(x\right)\eta_2=0,\\\label{e74} &\eta^{'}_4+\cos
\phi\left(x\right)\eta_2-(E-k) \eta_3-\sin \phi\left(x\right)\eta_1=0.
\end{align}

This set of equations for the form chosen for $\phi(x)$, which is kink, cannot be solved analytically. Thus, we use again an appropriate numerical method. Our purpose is to find the scattering solutions of the system, not the parity eigenstates. Therefore, we need to find the wavefunctions for the whole interval $(-\infty,+\infty)$. We solve this set as an initial value problem, using the so-called Runge-Kutta methods. In order to find the solutions with high accuracy, we use the Runge-Kutta-Fehlberg method of order 6. We take advantage of the simple exactly solvable model to determine the initial boundary values for solving the equations.
We already have all the solutions of this model, including the wavefunctions for the scattering process \cite{dr,levinsondr}. Since the form chosen for the pseudoscalar field is similar to kink as the spatial variable $x$ tends to infinity, we can use the values of the scattering wavefunctions of this simple model as the initial boundary conditions for our model. Therefore, we start at $x=+\infty$ with the values of the scattering wavefunctions for the exactly solvable model (after dropping the factor $\mathrm{e}^{i k x}$ and up to the normalization factor) and go backward in the $x$-interval to find the values of the $\eta_i$s for all the mesh-points of the interval $(-\infty,+\infty)$ by solving Eqs.\,(\ref{e7}-\ref{e74}). Figure (\ref{fig.1}) shows $\eta_i$s for the case with the parameters $\theta_0=\pi$, $\mu=10$, $k=3.0$ and $E=+\sqrt{k^2+M^2}$. This figure also shows $\eta_i$s of the simple exactly solvable model, for comparison. Note that the oscillations are less pronounced on the left for the kink model. This indicates that there is less ``reflection'' from the kink, as is also evident from the graph of $\rho(x)$ shown in Fig.\,(\ref{density}). It is worth mentioning that the kink is totally reflectionless for the elementary bosons within the $\lambda \phi^4$ theory.
\begin{widetext}
\begin{center}
\begin{figure}[th] \includegraphics[width=11.8cm]{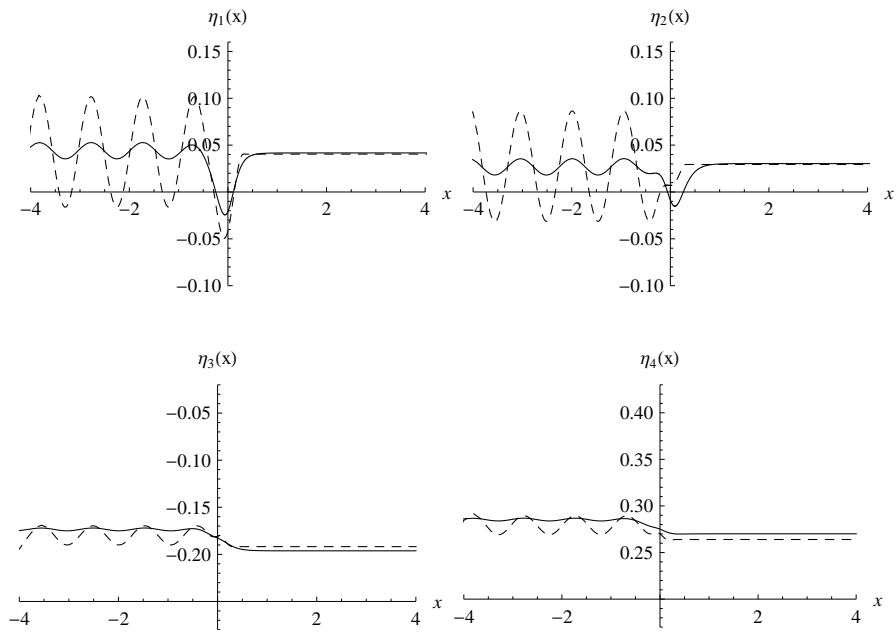}\caption{\label{fig.1} \small
   The graphs of $\eta_i(x)$s as functions of the spatial variable $x$, for the parameters $\theta_0=\pi$, $\mu=10$, $k=3.0$ and $E=+\sqrt{k^2+M^2}$. The solid and dashed lines show the normalized $\eta_i(x)$s for our model and the simple exactly solvable model, respectively.}
  \label{geometry}
\end{figure}
\end{center}
\end{widetext}
\begin{center}
\begin{figure}[th] \includegraphics[width=7.cm]{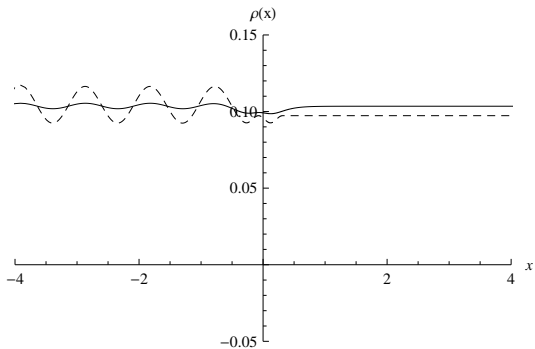}\caption{\label{density} \small
   The graph of $\rho(x)$ as a function of the spatial variable $x$, for the parameters $\theta_0=\pi$, $\mu=10$, $k=3.0$ and $E=+\sqrt{k^2+M^2}$. The solid and dashed lines show the $\rho(x)$ for our model and the simple exactly solvable model, respectively. The amplitude of oscillations on the left is proportional to the probability of ``reflection'' and the amplitude on the right is proportional to the ``transmission'' probability.}
  \label{geometry}
\end{figure}
\end{center}
\section{The calculation of the Casimir energy using the phase shift method}
 The Casimir energy for a system like ours, as is well-known, is given by the shift in the zero-point energies of fermionic modes due to the presence of the disturbance, and in general can be written in the following form
\begin{align}\label{e2}
E_{\mathrm{Casimir}}=&\int_{-\infty}^{+\infty}\text{d}x\int_{0}^{+\infty}\frac{\text{d}p}{2\pi}\sum\limits_{j=\pm}\left(-\sqrt{p^2+M^2}\right)\nu_p^{j\dag}\nu_p^j\nonumber\\
+&\int_{-\infty}^{+\infty}\text{d}x\sum\limits_{i}\left(E_{\text{bound}}^{i-}\right)
\chi_{2\text{b}_i}^\dag\chi_{2\text{b}_i}\nonumber\\
-&\int_{-\infty}^{+\infty}\text{d}x\int_{-\infty}^{+\infty}\frac{\text{d}k}{2\pi}\left(-\sqrt{k^2+M^2}\right)v_k^\dag
v_k\nonumber\\=-&\int_{0}^{+\infty}\text{d}k\sqrt{k^2+M^2}\left(\rho^{\mathrm{sea}}(k)-\rho_0^{\mathrm{sea}}(k)\right)\nonumber\\
+&\sum\limits_{i}E_{\text{bound}}^{i-}+\frac{M}{2}.
\end{align}
The first equality is the relation we derived in \cite{la} for the Casimir energy of a Fermi field in the presence of an arbitrary disturbance. The functions $\nu_p^j(x)$ and $v_k(x)$ are the normalized fermion wavefunctions for the negative continuum states in the presence and absence of the disturbance, respectively. The functions $\chi_{2\text{b}_i}(x)$ are the normalized fermion wavefunctions for the discrete bound states with negative energy and $E_{\text{bound}}^{i-}$ denote the energies of these negative bound states. In the last line of the above equation, the extra $M/2$ takes into account the contribution from the half-bound state of the fermion at $E=- M$ in the free case. The factor $(\rho^{\mathrm{sea}}(k)-\rho_0^{\mathrm{sea}}(k))$ is the difference between the density of continuum states with negative energy in the presence and absence of the pseudoscalar field.
\par In \cite{la} we concluded that for the simple exactly solvable model we can calculate the Casimir energy only from the negative states or only the positive states, or the average of all of the states and the results are exactly the same in all cases. Since all the symmetries of the model with kink are the same as the simple exactly solvable model, the aforementioned argument is also true for the present model.
\subsection{The phase shift and Levinson theorem}
The difference between the density of the continuum states in the free and interacting cases can be written in terms of the scattering phase shift in the following form
\begin{equation}\label{e10}
\rho(k)-\rho_0(k)=\frac{1}{\pi}\frac{\mathrm{d}}{\mathrm{d}k}\delta(k),
\end{equation}
where $\delta(k)=\delta_{\mathrm{sky}}(k)+\delta_{\mathrm{sea}}(k)$, i.e. $\delta(k)$ sums over the contributions from both positive and negative energies. This relation is also true for the sea and sky, separately.
Therefore, the second term in the relation of the Casimir energy, Eq.\,(\ref{e2}), can be written in terms of the phase shift, as follows
\begin{align}\label{e11}
-&\int_{0}^{+\infty}\text{d}k\sqrt{k^2+M^2}\left(\rho^{\mathrm{sea}}(k)-\rho_0^{\mathrm{sea}}(k)\right)\nonumber\\=&
-\int_{0}^{+\infty}\frac{\text{d}k}{\pi}\sqrt{k^2+M^2}\frac{\mathrm{d}}{\mathrm{d}k}\delta^{\mathrm{sea}}(k)\nonumber\\=&-\int_{0}^{+\infty}\frac{\text{d}k}{\pi}\sqrt{k^2+M^2}\frac{\mathrm{d}}{\mathrm{d}k}\left(\delta^{\mathrm{sea}}(k)-\delta^{\mathrm{sea}}(\infty)\right)\nonumber\\=&\int_{0}^{+\infty}\frac{\text{d}k}{\pi}\frac{k}{\sqrt{k^2+M^2}}\left(\delta^{\mathrm{sea}}(k)-\delta^{\mathrm{sea}}(\infty)\right)\nonumber\\
+&\frac{1}{\pi}M\left(\delta^{\mathrm{sea}}(0)-\delta^{\mathrm{sea}}(\infty)\right).
\end{align}
In the second equality we have just subtracted a zero term from the original one. For the last equality we have integrated the expression by parts, since the final expression is more convenient for the numerical analysis. Therefore, we can compute the second term in the expression of the Casimir energy using the phase shift.
\par Now, by comparing the coefficients of $\mathrm{e}^{i k x}$ on the left- and right-hand sides of the scattering region, we can obtain the scattering matrix element, which is related to the phase shift as $S(k)=\mathrm{e}^{i \delta(k)}$. We know that for $x\rightarrow -\infty$ the wavefunction would be a linear combination of $\mathrm{e}^{i k x}$ and $\mathrm{e}^{-i k x}$. Therefore, to obtain the coefficients of $\mathrm{e}^{i k x}$ on the left-hand side, we define the wavefunction of the far left as follows
\begin{equation}\label{e9}
 \left(\! \begin{array}{c}a_1+i a_2\\
a_3+i a_4\end{array}\! \right)\mathrm{e}^{-i k x}+ \left(\! \begin{array}{c}b_1+i b_2\\
b_3+i b_4\end{array}\! \right)\mathrm{e}^{i k x}=  \left(\! \begin{array}{c}\eta_1+i \eta_2\\
\eta_3+i \eta_4\end{array}\! \right)\mathrm{e}^{i k x},
\end{equation}
where $\eta_i$s on the right-hand side are obtained from the numerical method for $x\rightarrow -\infty$. Since this relation has been written for the left-hand side and far from the scattering region, all the elements $a_i$ and $b_i$ are constants, independent of the spatial variable $x$. The wavefunction in the left-hand side of Eq.\,(\ref{e9}) has to satisfy the first-order equations of motion (Eq.\,(\ref{e5})) when $x$ tends to minus infinity in this set of equations. Using the relation obtained from the substitution of this wavefunction into Eq.\,(\ref{e5}) and using Eq.\,(\ref{e9}), we can obtain all the constants $a_i$ and $b_i$. It is important to note that in the numerical computations we can only approach $x=-\infty$ or equivalently $X=-1$. This decreases the accuracy, when the slope of $\phi(x)$ at $x=0$ is very small, since in this case the region of variation of the pseudoscalar field becomes extended.
\par
\begin{center}
\begin{figure}[th] \includegraphics[width=6.8cm]{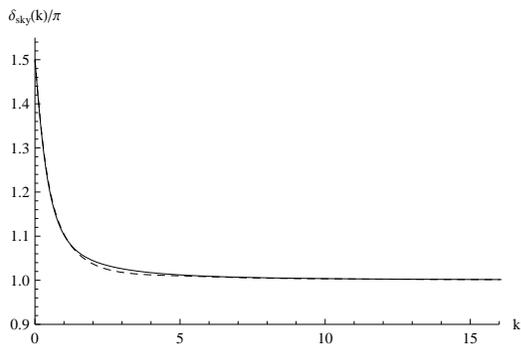}\caption{\label{fig.11} \small
   The graphical representation of $\delta_{\mathrm{sky}}(k)$, for $\mu=10$ and $\theta_0=\pi$. The results for the models with kink and the simple exactly solvable model are shown by the solid and dashed lines, respectively. For these parameters we can see from Fig.\,(\ref{bnd}) that two levels have exited from the sky and since $N^0_{\mathrm{t},\mathrm{sky}}=1$ at $E=+1.0$, we expect $\delta_{\mathrm{sky}}(0)/{\pi}=3/2$.}
  \label{geometry}
\end{figure}
\end{center}

\begin{center}
\begin{figure}[th] \includegraphics[width=6.8cm]{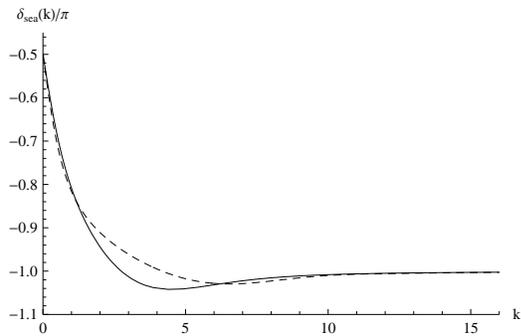}\caption{\label{fig.12} \small
    The graphical representation of $\delta_{\mathrm{sea}}(k)$, for $\mu=10$ and $\theta_0=\pi$. The results for the models with kink and the simple exactly solvable model are shown by the solid and dashed lines, respectively. For these parameters we can see from Fig.\,(\ref{bnd}) that no level has entered the sea and since $N^0_{\mathrm{t},\mathrm{sea}}=1$ at $E=-1.0$, we expect $\delta_{\mathrm{sea}}(0)/{\pi}=-1/2$.}
  \label{geometry}
\end{figure}
\end{center}

 Now we check the consistency of the resulting phase shifts with the Levinson theorem. The weak form of this theorem for the Dirac equation is as follows \cite{levinsondr}
\begin{eqnarray}\label{levweak}
\Delta \delta  &\equiv&[\delta_{\mathrm{sky}}(0)-\delta_{\mathrm{sky}}(\infty)]+[\delta_{\mathrm{sea}}(0)-\delta_{\mathrm{sea}}(\infty)]\nonumber\\
&=&\left(N+\frac{N_\mathrm{t}}{2}-\frac{N_\mathrm{t}^0}{2}\right)\pi,
\end{eqnarray}
where $N$ is the total number of bound states, including positive and negative ones, $N_t$ the total number of the threshold bound states at the given strength of the potential, and $N_t^0$ the number of bound states at zero strength of the potential, i.e. the free Dirac case. The strong form of the Levinson theorem relates the value of the phase shift at each boundary of the continua to the number of levels which have crossed those boundaries, in the process of building up the disturbance. This form of theorem can be expressed in the following form for $k=0$
\begin{equation}\label{levstrong}
 \delta(0)=\left(N_{\mathrm{exit}}-N_{\mathrm{enter}}\right)\pi.
\end{equation}
This relation holds for each of the continua, separately. For each continuum $N_{\mathrm{exit}}$ ($N_{\mathrm{enter}}$) is the number of the bound states that exit (enter) that continuum from that boundary ($E=+1.0$ or $E=-1.0$) as the strength of the potential is increased from zero to its final finite value. In this equation the threshold bound states, mentioned above, should be included in $N_{\mathrm{exit}}$ and $N_{\mathrm{enter}}$, with the coefficient $1/2$. Moreover, the strong form of the Levinson theorem for $k=\infty$ can be written in the following form
\begin{equation}\label{levstrong2}
 \delta(\infty)=\left(N_{\mathrm{enter}}-N_{\mathrm{exit}}\right)\pi.
\end{equation}
This relation also holds for each continuum, separately. For each continuum $N_{\mathrm{exit}}$ ($N_{\mathrm{enter}}$) is the number of the bound states that exit (enter) that continuum from $k=\infty$ ($E=+\infty$ or $E=-\infty$) as the strength of the potential is increased from zero to its final finite value. As an example of our results, in Figs.\,(\ref{fig.11}) and (\ref{fig.12}) we plot the phase shift for our system as a function of $k$, for the parameters $\theta_0=\pi$ and $\mu=10$. Figure (\ref{fig.11}) shows the $\delta_{\mathrm{sky}}(k)$, i.e. the phase shift for the states with the positive energy $+\sqrt{k^2+M^2}$ and Fig.\,(\ref{fig.12}) shows $\delta_{\mathrm{sea}}(k)$, i.e. the phase shift for the states with the negative energy $-\sqrt{k^2+M^2}$. In both figures we also show the phase shift of the simple exactly solvable model with the same parameters, for comparison.
 It is easy to check that the phase shifts depicted in these figures are consistent with both the weak and strong forms of the Levinson theorem. In particular $\delta(E=\pm \infty)=\pm \theta_0$, which is consistent with the results of the adiabatic method of Goldstone and Wilczek \cite{goldstone}.

\subsection{The Casimir energy}
\begin{center}
\begin{figure}[th] \includegraphics[width=6.9cm]{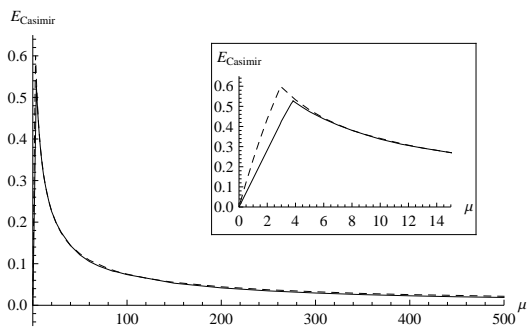}\caption{\label{fig.13} \small
   The graphical representation of the Casimir energy as a function of $\mu$, the scale of variation of the soliton, at $\theta_0=\pi$. The solid and dashed lines show the results for the model with the kink and the simple exactly solvable model, respectively. In the box we focus on the small values of $\mu$ to show the details of the maximum and the differences between the results of two models.}
  \label{geometry}
\end{figure}
\end{center}

Now we can calculate the Casimir energy using Eq.\,(\ref{e2}) and show the results in some figures. In Fig.\,(\ref{fig.13}) we plot the Casimir energy as a function of $\mu$ at $\theta_0=\pi$, i.e. a soliton with winding number 1, for our model and for the simple exactly solvable model. As can be seen in both cases, there is a sharp maximum occurring when the bound energy level crosses the line of $E=0$. This crossing happens for a larger value of $\mu$ for our model, based on the bound energy levels shown in the left graph of Fig.\,(\ref{bnd}). For the model with kink the bound energy level crosses the line of $E=0$ at $\mu \approx 3.821$, while for the simple exactly solvable model this crossing occurs at a lower value of the slope, i.e. $\mu\approx 2.957$. Also, the value of the Casimir energy is lower in the case of kink. The largest difference between the graphs of these two models occurs around the maximum, as is shown in the zoomed box of this figure. However, when $\mu\rightarrow 0$ or $\mu\rightarrow \infty$, both graphs tend to the same values. In both models the Casimir energy reaches the expectable value of zero when the slope of the soliton at $x=0$ decreases to zero, i.e. when the vacuum energy approaches that of the trivial vacuum, despite the residual non-trivial boundary conditions. Also, both graphs have the same limit when the slope of the soliton tends to infinity. This limit is zero at $\theta_0=n\pi$, i.e. when we have a proper soliton with winding number $n$. However, for other values of $\theta_0$ the Casimir energy is in general non-zero when $\mu\rightarrow \infty$ in both models (see Eq.\,(3.9) in \cite{la}).
\par
\begin{center}
\begin{figure}[th] \includegraphics[width=7.1cm]{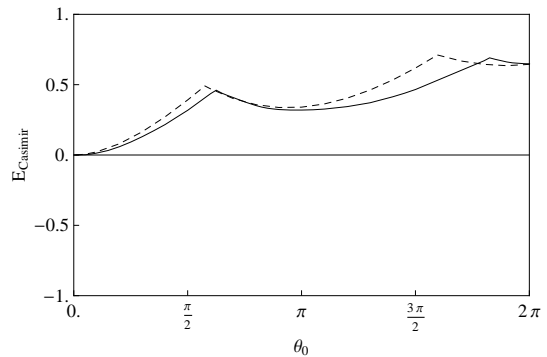}\caption{\label{casth} \small
   The Casimir energy as a function of $\theta_0$, at $\mu=10$. The solid and dashed lines show the results for the model with the kink and the simple exactly solvable model, respectively. The cusps occur when bound state energy levels cross $E=0$. The bound energy levels cross $E=0$ at $\theta_0\approx 0.601\pi$ and $\theta_0\approx 1.801\pi$, for the model with kink, and at $\theta_0\approx 0.576\pi$ and $\theta_0\approx 1.596\pi$, for the simple exactly solvable model, in the interval $0\leqslant\theta_0\leqslant 2\pi$.}
  \label{geometry}
\end{figure}
\end{center}

 In Fig.\,(\ref{casth}) we present the Casimir energy as a function of $\theta_0$ for $\mu=10$, by the solid and dashed lines for the model with kink and the simple exactly solvable model, respectively. As can be seen, the Casimir energy is, on the average, an increasing function of $\theta_0$ for both models and there are two mild cusps in each graph. Comparing these graphs with the right graph of Fig.\,(\ref{bnd}), we conclude that these cusps occur when the bound energy levels cross the line of $E=0$.
\section{Stability of the solutions}
\begin{center}
\begin{figure}[th] \includegraphics[width=6.6cm]{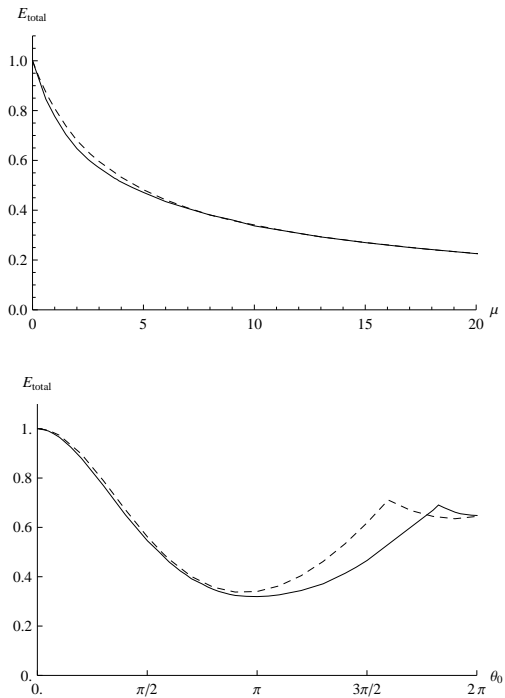}\caption{\label{stability} \small
  The upper graph shows the total energy (the sum of the energy of a valence fermion and the Casimir energy) as a function of $\mu$ when $\theta_0=\pi$. The lower graph shows the total energy as a function of $\theta_0$ for the slope $\mu=10$. In both graphs the solid and dashed lines represent the total energy for the model with kink and the simple exactly solvable model, respectively. Notice that the system attains its lowest energy at $\theta_0=\pi$.}
  \label{geometry}
\end{figure}
\end{center}

Finally, we consider the effect of the Casimir energy on the total energy of a system consisting of a valence fermion in the ground state. The total energy for such a system consists of the Casimir energy and the energy of the valence fermion. Note that the energy of the valence fermion should not be added when this energy is negative, since it has already been taken into account in the Casimir energy. The total energy is shown in Fig.\,(\ref{stability}). The upper graph shows this energy as a function of the slope of the pseudoscalar field ($\mu$) when $\theta_0=\pi$ and the lower graph shows this energy as a function of the value of the pseudoscalar field at infinity ($\theta_0$) when $\mu=10$. In both graphs the solid and dashed lines represent the total energy for the model with kink and the simple exactly solvable model, respectively. As can be seen in the lower graph, for both models there is a minimum occurring at $\theta_0\approx \pi$, which corresponds to a soliton with winding number one. This means that not only this configuration is energetically favorable, but also it is stable against small fluctuations in the parameters of the background field when this field is a soliton with a proper winding number, as expected. Notice that there is no minimum in the upper graph.

\section{Conclusion}
In this paper we compute the Casimir energy for a Fermi field chirally coupled to a pseudoscalar field chosen to be prescribed and in the form of the kink. The equations of motion for this system are not exactly solvable. Using the Runge-Kutta-Fehlberg method of order 6 and taking the scattering solutions of an exactly solvable model containing a fermion and a pseudoscalar field in the simple form of a piece-wise linear function, as the initial condition, we find the scattering wavefunctions for the fermion in the system with the kink. Comparing the graphs of the scattering probability for the system with the kink and the one with the simple background field shows that the kink is almost reflectionless for the fermion, within the context of our model. Then, comparing the values of these wavefunctions at the boundaries of the $x$ interval, i.e. $x\rightarrow \pm \infty$, we obtain the phase shifts for the fermion's wavefunctions and plot them for the specific set of parameters of the model. We check the consistency of the resulting phase shifts with the weak and strong forms of the Levinson theorem and conclude that this theorem is completely valid for our results. Then, using the relation between the derivative of the phase shift and the difference between the density of states in the presence and absence of the kink, we calculate the Casimir energy of our system and depict this energy as a function of the parameters determining the kink, i.e. its scale of variations ($\mu$) and its value at infinity ($\theta_0$). We show that in the graph of the Casimir energy as a function of $\mu$ there is a sharp maximum occurring when the bound energy level of the fermion crosses the line of $E=0$. Since this crossing occurs at a larger value of $\mu$ for the model with kink as compared to the simple exactly solvable model, the maximum of the Casimir energy for the model with kink occurs at a larger value of $\mu$. Moreover, this graph shows the expectable results at the limits $\mu\rightarrow 0$ and $\mu\rightarrow \infty$. For $\theta_0=\pi$ which describes a kink with winding number one, both limits are zero. The Casimir energy is in general an increasing function of $\theta_0$ and is always positive and has a cusp whenever there is a zero fermionic mode. The first few cusps are local maxima. Finally, we compute the total energy of a system consisting of a valence fermion in the ground state. This energy includes the Casimir energy of the system and the energy of the valence fermion. We conclude that there is no preferable $\mu$ for the system consisting of the background field with winding number one. However, considering the effect of the changing the $\theta_0$ for fixed $\mu$, the system is more stable for the soliton with winding number one.

\begin{acknowledgments} \label{Calculation}
We would like to thank the research office of the Shahid Beheshti University for financial support.
\end{acknowledgments}


\begin{thebibliography}{9}
\bibitem{casimir1}
H.B.G. Casimir, Proc. Kon. Aa. Wet. \textbf{51}, 793 (1948).
\bibitem{casimir2}
H.B.G. Casimir and D. Polder, Phys. Rev. \textbf{73}, 360 (1948).
\bibitem{plate1}
V.B. Bezerra, G. Bimonte, G.L. Klimchitskaya, V.M. Mostepanenko and C. Romero, Eur. Phys. J. C \textbf{52}, 701 (2007).
\bibitem{plate2}
R. Moazzemi, M. Namdar and S.S. Gousheh, JHEP \textbf{09}, 029 (2007).
\bibitem{plate3}
R. Moazzemi and S.S. Gousheh, Phys. Lett. \textbf{B 658}, 255 (2008).
\bibitem{plate4}
S.S. Gousheh, R. Moazzemi and M.A. Valuyan, Phys. Lett. \textbf{B 681}, 477 (2009).
\bibitem{plate5}
H. Cheng, Phys. Rev. \textbf{D 82}, 045005 (2010).
\bibitem{deraadcylinder}
L.L. DeRaad and K.A. Milton, Ann. Phys. (N.Y.) \textbf{136}, 229 (1981).
\bibitem{cylinder1}
F.D. Mazzitelli, M.J. Sanchez, N.N. Scoccola and J. von Stecher, Phys. Rev. \textbf{A 67}, 013807 (2003).
\bibitem{cylinder2}
D.A.R. Dalvit, F.C. Lombardo, F.D. Mazzitelli and R. Onofrio, Europhys. Lett. \textbf{67}, 517 (2004).
\bibitem{cylinder3}
P.A.M. Neto, J. Opt. B: Quantum Semiclass. Opt. \textbf{7}, s86 (2005).
\bibitem{cylinder4}
D.A.R. Dalvit, F.C. Lombardo, F.D. Mazzitelli and R. Onofrio, Phys. Rev. \textbf{A 74}, 020101 (2006).
\bibitem{cylinder5}
E.K. Abalo, K.A. Milton and L. Kaplan, Phys. Rev. \textbf{D 82}, 125007 (2010).
\bibitem{boyersphere}
T.H. Boyer, Phys. Rev. \textbf{174}, 1764 (1968).
\bibitem{sphere1}
R. Balian and B. Duplantier, Ann. Phys. (N.Y.) \textbf{112}, 165 (1978).
\bibitem{sphere2}
K.A. Milton, L.L. DeRaad and J. Schwinger, Ann. Phys. (N.Y.) \textbf{115}, 388 (1978).
\bibitem{sphere3}
C.M. Bender and K.A. Milton, Phys. Rev. \textbf{D 50}, 6547 (1994).
\bibitem{sphere4}
M. Bordag, E. Elizalde, K. Kirsten and S. Leseduarte, Phys. Rev. \textbf{D 56}, 4896 (1997).
\bibitem{othergeometry1}
W. Lukosz, Physica \textbf{56}, 109 (1971).
\bibitem{othergeometry2}
J.R. Ruggiero, A. Villani and A.H. Zimerman, J. Phys. A: Math. Gen. \textbf{13}, 761 (1980).
\bibitem{othergeometry3}
S. Hacyan, R. Jauregui and C. Villarreal, Phys. Rev. \textbf{A 47}, 4204 (1993).
\bibitem{othergeometry4}
G.J. Maclay, Phys. Rev. \textbf{A 61}, 052110 (2000).
\bibitem{othergeometry5}
X. Li and X. Zhai, J. Phys. A: Math. Gen. \textbf{34}, 11053 (2001).
\bibitem{othergeometry6}
H. Cheng, J. Phys. A: Math. Gen. \textbf{35}, 2205 (2002).
\bibitem{othergeometry7}
M.A. Valuyan, R. Moazzemi and S.S. Gousheh, J. Phys. B: At. Mol. Opt. Phys. \textbf{41}, 145502 (9pp) (2008).
\bibitem{othergeometry8}
A. Seyedzahedi, R. Saghian and S.S. Gousheh, Phys. Rev. \textbf{A 82}, 032517 (2010).
\bibitem{heat1}
T.P. Branson and P.B. Gilkey, Commun. Partial differential Eqs.
\textbf{15}, 245 (1990).
\bibitem{heat2}
M. Bordag and K. Kiresten, Int. J. Mod. Phys. \textbf{A 17}, 813
(2002).
\bibitem{green}
K.A. Milton, L.L. Deraad and J. Schwinger, Ann. Phys. (N.Y.) \textbf{115}, 388 (1978).
\bibitem{modenumber1}
I.H. Brevik, V.V. Nesterenko and I.G. Pirozhenko, J. Phys. \textbf{A
31}, 8661 (1998).
\bibitem{modenumber2}
K.A. Milton, A.V. Nesterenko and V.V. Nesterenko, Phys. Rev.
\textbf{D 59}, 105009 (1999).
\bibitem{modenumber3}
V.V. Nesterenko and I.G. Pirozhenko, J. Math. Phys. \textbf{41}, 4521 (2000).
\bibitem{modenumber4}
A. Romeo and K.A. Milton, Phys. Lett. \textbf{B 621}, 309 (2005).
\bibitem{scatter}
R. Balian and B. Duplantier, Ann. Phys. (N.Y.) \textbf{112}, 165
(1978).
\bibitem{sparnaay}
M.J. Sparnaay, Physica \textbf{24}, 751 (1958).
\bibitem{Lamorea1}
S.K. Lamoreaux, Phys. Rev. Lett. \textbf{78}, 5 (1997).
\bibitem{Lamorea2}
 S.K. Lamoreaux, Phys. Rev. Lett. \textbf{81}, 5475(E) (1998).
\bibitem{exper1}
U. Mohideen and A. Roy, Phys. Rev. Lett. \textbf{81}, 4549 (1998).
\bibitem{exper2}
A. Roy and U. Mohideen, Phys. Rev. Lett. \textbf{82}, 4380 (1999).
\bibitem{exper3}
Th. Ederth, Phys. Rev. \textbf{A 62}, 062104 (2000).
\bibitem{exper4}
G. Bressi, G. Carugno, R. Onofrio and G. Ruoso, Phys. Rev. Lett. \textbf{88}, 041804 (2002).
\bibitem{exper5}
R.S. Decca, D. L\'{o}pez, H.B. Chan, E. Fischbach, D.E. Krause and C.R. Jamell, Phys. Rev. Lett.
\textbf{94}, 240401 (2005).
\bibitem{exper6}
R.S. Decca, D. L\'{o}pez, E. Fischbach, G.L. Klimchitskaya, D.E. Krause and V.M. Mostepanenko, Annals of Physics (N.Y.) \textbf{318}, 37 (2005).
\bibitem{exper7}
G. Bimonte, D. Born, E. Calloni, G. Esposito, U. Huebner, E.
Il'ichev,
 L. Rosa, F.Tafuri and R. Vaglio, J. Phys. A: Math. Theor. \textbf{41}, 164023 (8pp) (2008).
\bibitem{dehghan}
Z. Dehghan and S.S. Gousheh, IJMPA \textbf{27}, 1250093 (2012).
 \bibitem{dashen}
R.F. Dashen, B. Hasslacher and A. Neveu, Phys. Rev. \textbf{D 10}, 4130 (1974).
\bibitem{solitonmass1}
J.L. Gervais and A. Neveu(eds.), Phys. Rept. \textbf{23}, 237 (1976).
\bibitem{solitonmass2}
H.J. Vega, Nucl. Phys. \textbf{B 115}, 411 (1976).
\bibitem{solitonmass3}
J. Verwaest, Nucl. Phys. \textbf{B 123}, 100 (1977).
\bibitem{solitonmass4}
L.D. Faddeev and V.E. Korepin, Phys. Rept. \textbf{42}, 1 (1978).
\bibitem{solitonmass5}
R. Rajaraman, \textit{Solitons and Instantons: An Introduction to Solitons and Instantons in Quantum Field Theory} (North-Holland, Amsterdam, 1982).
\bibitem{solitonmass6}
M. Bordag, A.S. Goldhaber, P. van Nieuwenhuizen and D. Vassilevich, Phys. Rev. \textbf{D 66}, 125014 (2002).
\bibitem{solitonmass7}
A. Rebhan, P. van Nieuwenhuizen and R. Wimmer, New J. Phys. \textbf{4}, 31 (2002).
\bibitem{solitonmass8}
G. Mussardo, V. Riva and G. Sotkov, Nucl. Phys. \textbf{B 699}, 545 (2004).
\bibitem{heavy1}
J.A. Bagger and S.G. Naculich, Phys. Rev. Lett. \textbf{67}, 2252
(1991); Phys. Rev. \textbf{D 45}, 1395 (1992).
\bibitem{heavy2}
S.G. Naculich, Phys. Rev. \textbf{D 46}, 5487 (1992).
\bibitem{heavy3}
E. Farhi, N. Graham, R.L. Jaffe and H. Wiegel, Nucl. Phys. \textbf{B 585}, 443 (2000).
\bibitem{heavy4}
E. Farhi, N. Graham, R.L. Jaffe and H. Wiegel, Phys. Lett. \textbf{B 475}, 335 (2000).
\bibitem{super1}
A. Rebhan and P. van Nieuwenhuizen, Nucl. Phys. \textbf{B 508}, 449 (1997).
\bibitem{super2}
A.S. Goldhaber, A. Litvintsev and P. van Nieuwenhuizen, Phys. Rev. \textbf{D 67}, 105021 (2003).
\bibitem{super3}
A. Rebhan, P. van Nieuwenhuizen and R. Wimmer, Nucl. Phys. \textbf{B 679}, 382 (2004).
\bibitem{la}
L. Shahkarami, A. Mohammadi and S.S. Gousheh, JHEP \textbf{11}, 140
(2011).
\bibitem{dr}
S.S. Gousheh and R. L\'{o}pez-Mobilia, Nucl. Phys. \textbf{B 428},
189 (1994).
\bibitem{leila}
L. Shahkarami and S.S. Gousheh,JHEP \textbf{06}, 116 (2011).
\bibitem{levinsondr}
S.S. Gousheh, Phys. Rev. \textbf{A 65}, 032719 (2002).
\bibitem{jackiw}
R. Jackiw and C. Rebbi, Phys. Rev. \textbf{D 13}, 3398
(1976).
\bibitem{goldstone}
J. Goldstone and F. Wilczek, Phys. Rev. Lett. \textbf{47}, 986 (1981).
\bibitem{mackenzie}
R. MacKenzie and F. Wilczek, Phys. Rev. \textbf{D 30}, 2194 (1984).
\end{thebibliography}
 \end{document}